\documentclass[epj]{svjour}
\usepackage{latexsym}
\usepackage{graphics}
\usepackage{graphicx}
\usepackage{amsfonts,mathbbol,euscript,mathrsfs}
\usepackage{latexsym,amssymb}
\usepackage{rotating,lscape}
\usepackage{fancyhdr}
\usepackage{tocbibind}
\usepackage{colortbl}
\usepackage{multirow}
\usepackage{textcomp}
\usepackage{color}
\usepackage{url}
\usepackage{psfrag}
\usepackage{subfigure}
\usepackage{sidecap}
\usepackage{makeidx}
\usepackage[intoc,prefix]{nomencl}
\usepackage{ifthen}
\usepackage{color}
\newcommand{\Rb}{$^{87}$Rb}

\newcommand{\Cs}{$^{133}$Cs}

\newcommand{\mF}{$m_{{F}}$}
\newcommand{\etal}{{\it et al.\ }}

\newcommand{\mFminus}{$\left|F=1,~m_{F}=-1\right\rangle$}
\newcommand{\abbmFplus}{$\left|1,+1\right\rangle$}
\newcommand{\abbmFzero}{$\left|1,0\right\rangle$}
\newcommand{\abbmFminus}{$\left|1,-1\right\rangle$}
\begin{document}
\title{Bose-Einstein condensation of \Rb\ in a levitated crossed dipole trap}
\author{D. L. Jenkin \and D. J. McCarron \and M. P. K{\"{o}}ppinger \and H.-W. Cho \and S. A. Hopkins \and S. L. Cornish}
\institute{Department of Physics, Durham University, Rochester Building, Science Laboratories,  South Road, Durham. DH1 3LE, United Kingdom}
\date{Received: date / Revised version: date}
\abstract{We report an apparatus and method capable of producing  Bose-Einstein condensates (BECs) of $\sim1\times$10$^{6}$ \Rb\ atoms, and ultimately designed for sympathetic cooling of \Cs\ and the creation of ultracold RbCs molecules. The method combines several elements: i) the large recapture of a magnetic quadrupole trap from a magneto-optical trap, ii) efficient forced RF evaporation in such a magnetic trap, iii) the gain in phase-space density obtained when loading the magnetically trapped atoms into a far red-detuned optical dipole trap and iv) efficient evaporation to BEC within the dipole trap. We demonstrate that the system is capable of sympathetically cooling the \mFminus\ and \abbmFzero\ sublevels with \abbmFplus\ atoms. Finally we discuss the applicability of the method to sympathetic cooling of \Cs\ with \Rb.}

\authorrunning{D. L. Jenkin \etal}
\titlerunning{Bose-Einstein condensation of \Rb\ in a levitated crossed dipole trap}
\maketitle
\section{\label{chap:Intro}Introduction}

The field of matterwave optics has progressed enormously since the first observations
of Bose-Einstein condensation (BEC) \cite{Anderson:OOBECIADAV,Davis:BECIAGOSA,Bradley:EOBECIAAGWAI} in 1995, spawning sub-fields such as atom chips, BEC in optical lattices, rotating BECs and matterwave interferometry. More recently the production of mixtures of two or more ultracold atomic gases has opened up an exciting field of rich physics; see recent reviews \cite{Chin:FRIUG,Carr:CAUMSTAA,Lahaye:TPODBQG,Bloch:MBPWUG}. Mixtures of various isotopic combinations of Li, Na, K, Rb and Cs have been prepared and studied (see Table IV of \cite{Chin:FRIUG}). Such mixtures open up the intriguing possibility of creating ultracold, dipolar, heteronuclear molecules, which possess long range dipole-dipole interactions that will allow the construction of versatile quantum simulators and quantum information processors \cite{Chin:FRIUG,Carr:CAUMSTAA,Lahaye:TPODBQG,Bloch:MBPWUG}, as well as having applications in precision measurement and studies of condensate dynamics. Mixtures also play an important technical role in the sympathetic cooling of `difficult' bosonic species such as {$^{85}$Rb} \cite{Papp:TMIATSBEC,Altin:85RTIBECM} {$^{41}$K} \cite{Thalhammer:DSBECWTII}, \Cs\ \cite{Thomas:SECTBECOAMTCG} and all fermions \cite{Ketterle:MPAUUFG} owing to the suppression of s-wave scattering for fermions. Sympathetic cooling is also expected to play an important role in many newly proposed schemes that will connect direct molecular cooling methods such as supersonic expansion, buffer-gas cooling and Stark deceleration with methods from the field of laser cooling \cite{Faraday:FD}.

For successful sympathetic cooling, there are two requirements. Firstly, it must be possible to collect a large number of the refrigerant species and secondly, it must be possible to evaporate predominantly the refrigerant species, whilst maintaining good thermal contact with the species being sympathetically cooled. These requirements are met in typical experiments by collecting a large number of atoms in a magneto-optical trap (MOT) and then evaporating the refrigerant after transfer to either a magnetic or optical dipole trap. Each of the latter two types of trap have advantages and disadvantages. For example, the simplest form of magnetic trap, the quadrupole trap, is easy to implement and can capture a large fraction of the atoms collected in the MOT. Moreover, its linear potential makes for efficient evaporation \cite{Ketterle:ECOTA} at constant stiffness. However only weak-field seeking states may be trapped, and the field zero at the trap centre leads to loss via Majorana spin flips at an increasing rate as the atoms are cooled. On the other hand, the dipole trap can confine all magnetic states, allows good optical access (as no coils are involved) and the magnetic field becomes a free parameter for Stern-Gerlach experiments or those using Feshbach resonances \cite{Chin:FRIUG}. The primary disadvantage of dipole traps is their small volume and low trap depth that necessitate prior cooling before loading.

Hybrid magnetic/dipole traps have been demonstrated \cite{Davis:BECIAGOSA,Hoang:CIACHOAMT,Hung:AECOAIBECIOT,Lin:RPO87RBECIACMAOP} and, in general, possess extra degrees of freedom for atomic manipulation compared to a single-natured trap. Prominent examples include i) the transfer from a pure magnetic trap to a hybrid trap that exploits the so-called `dimple' effect \cite{Lin:RPO87RBECIACMAOP,Kraemer:OPOACBEC,Stamper-Kurn:RFOABEC,KleineBuning:ASGCAL} resulting in increased phase-space density (PSD) and  ii) the use of the gravito-magnetic `tilting' technique \cite{Hung:AECOAIBECIOT} that allows efficient evaporation in a crossed dipole trap where the trap frequencies remain approximately constant as the well depth is lowered.

In this work, we report a novel method and apparatus for sympathetic cooling of mixtures which involves a successful combination of the schemes i) and ii) above. Our experiment is intended for cooling of `difficult' \Cs\ by a refrigerant of `easy' \Rb\ with the ultimate aim of creating ultracold RbCs molecules, but the method has more general applicability to mixtures of other species.

\begin{figure}
\centering
\includegraphics[width=\columnwidth]{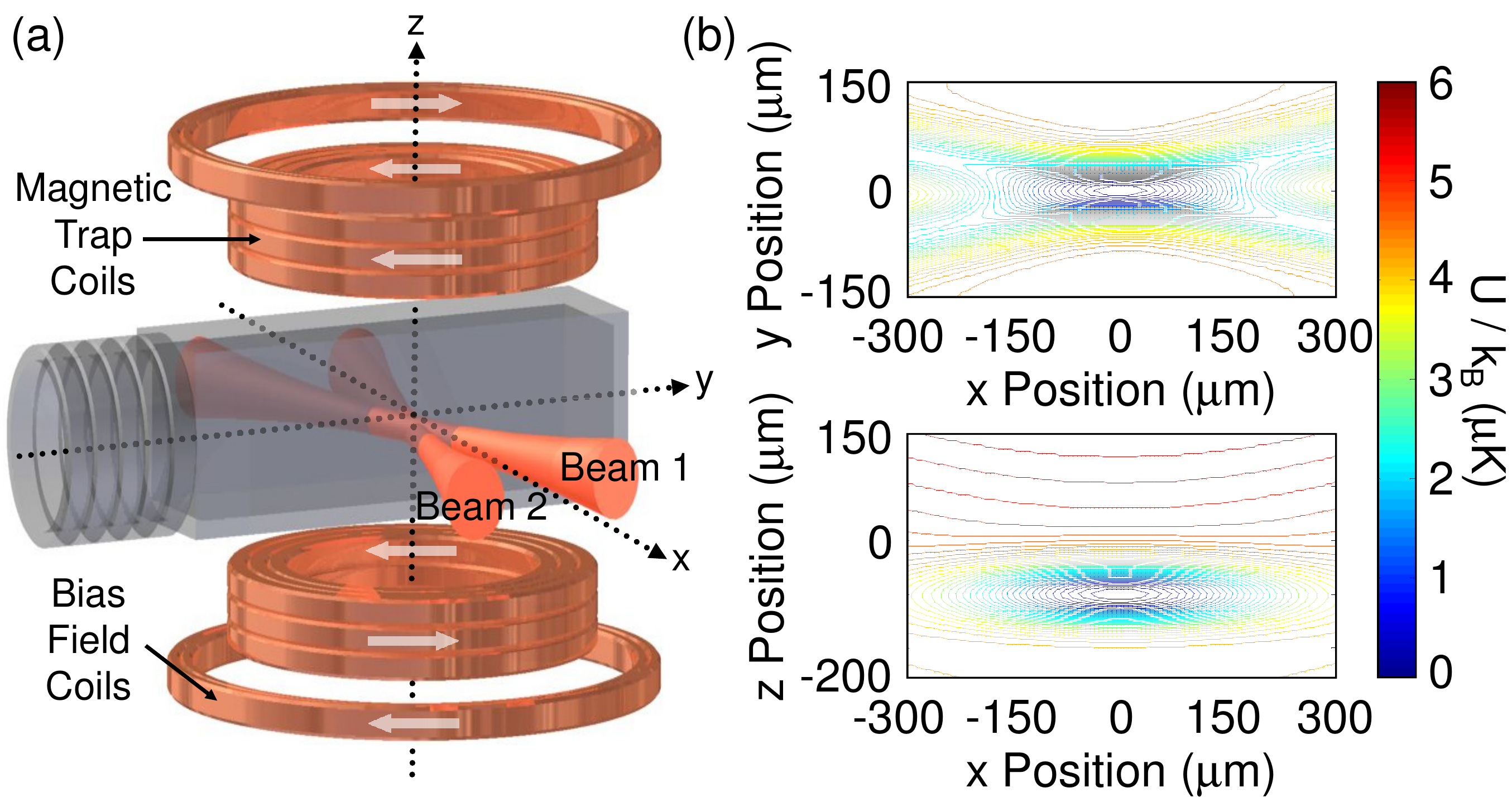}
\caption{The levitated crossed dipole trap. (a) Schematic of the trap geometry showing the coils used to generate the magnetic potential and the intersection of the two dipole beams within the ultra high vacuum (UHV) glass cell to create the optical potential. (b) Typical contour plots of the resulting hybrid trapping potential including gravity for \Rb\ in the \abbmFplus\ state. In this example both 150~mW beams are focussed to $\sim60~\mu$m, the magnetic field gradient is 29~G~cm$^{-1}$ and the bias field is 22.8~G along $z$.}
\label{fig:Diagram}%
\end{figure}

The basic apparatus for the sympathetic cooling, illustrated in Figure~\ref{fig:Diagram}~(a) consists of a combination of a crossed dipole trap positioned directly below the centre of a magnetic quadrupole trap. A second set of `bias' coils produces a variable magnetic bias field at the dipole trap and also shifts the position of the quadrupole field zero along the vertical axis in order to levitate the atoms. Figure~\ref{fig:Diagram} (b) shows typical contour plots of the total potential (optical + magnetic + gravitational). These plots are made for a specific configuration, the `levitated trap', to be discussed later in Sections~\ref{sec:LevTrap}~and~\ref{chap:Expt}, but our intention at this stage is to show the generic shape. 
 
The overall experimental routine is as follows. \Rb\ atoms are loaded into a standard six-beam MOT using a separate pyramid MOT as a cold atom source \cite{Harris:MTOACRBAM}. The atoms are then optically pumped into the \abbmFminus\ state and loaded into a magnetic quadrupole trap. Forced RF evaporation then reduces the temperature and increases the PSD prior to  loading the gas into the dipole trap. To load the dipole trap the magnetic field gradient is decreased to a value close to that which cancels gravity. At this stage the hybrid trap is in a configuration that we refer to as the `loading trap'. The atoms are then transferred into the absolute ground state \abbmFplus\ via RF driven rapid adiabatic passage \cite{Bergmann:CPTAQSOAAM}. Evaporative cooling is then carried out by reducing the dipole beam powers. 

The trapping potential of the `loading trap' and the `levitated trap' are described in detail in Section~\ref{chap:TrapPot}. In Section~\ref{chap:Expt} we describe other aspects of the experimental setup and the methods for optimally transferring atoms initially to the `loading trap' and subsequently to the levitated crossed dipole trap. Section~\ref{chap:Evap} presents results on cooling \Rb\ to degeneracy and sympathetic cooling of the different \mF\ sublevels. Finally Section~\ref{chap:Out} is a summary and outlook, and in particular discusses the applicability of the apparatus to trap and sympathetically cool \Cs.

\section{\label{chap:TrapPot}The trap potentials}

In this section we describe in detail two important configurations of the hybrid trap, namely the `loading trap' (Section~\ref{sec:LoadTrap}), which is optimised to capture a large number of atoms at the highest possible PSD, and the `levitated trap' (Section~\ref{sec:LevTrap}) which is optimised for the last stages of evaporation. The hybrid trap is composed of the sum of optical, magnetic and gravitational potentials. The optical potential is produced at the intersection of two horizontal laser beams with waists of $\sim60~\mu$m and wavelength 1550 nm, derived from a 30 Watt IPG ELR-30-LP-SF fibre laser. The beam directions are separated by an angle of 22$^{\circ}$ in the horizontal plane and the waists are positioned $\sim80~\mu$m (approximately a beam waist) below the centre of the unbiased quadrupole potential. The beams' frequencies are each shifted oppositely by $\pm$50~MHz using acousto-optic modulators (AOMs) to avoid standing wave effects. The intensity of each beam is monitored on a photodiode and servomechnically stabilised using the AOMs, thereby permitting precise control of the beam powers.

The optical potential is given by \cite{Grimm:ODTFNA}
\begin{equation}
U(\vec{r})~=~-\frac{1}{2\epsilon_{0}c}\Re(\alpha)I(\vec{r})~\approx~\frac{3\pi c^{2}}{2\omega^{3}_{0}}\frac{\Gamma}{\Delta}I(\vec{r}),
\label{eq:DipPot}
\end{equation}
where $\alpha$ is the complex polarisability, $\omega_{0}$ is the dominant optical transition frequency, $\Gamma$ is the spontaneous decay rate of the excited level, $\Delta$ is the detuning of the light from the atomic resonance and $I(\vec{r})$ is the intensity of the beam at position $\vec{r}$. The approximation is valid in the limit of large detuning and small saturation.

The magnetic field gradient is produced by a pair of coils wound in the anti-Helmholtz configuration and a magnetic bias field is produced by a coil pair wound in the Helmholtz configuration (see Figure~\ref{fig:Diagram}). We calculate the magnetic potential precisely by integrating the Biot-Savart law over the coil windings. For low-field seekers in a trapped state the magnetic potential adds to the optical potential, however for high-field seekers the magnetic potential reduces the overall trapping potential.

\subsection{\label{sec:LoadTrap}The loading trap}

\begin{figure}
\centering
\includegraphics[width=\columnwidth]{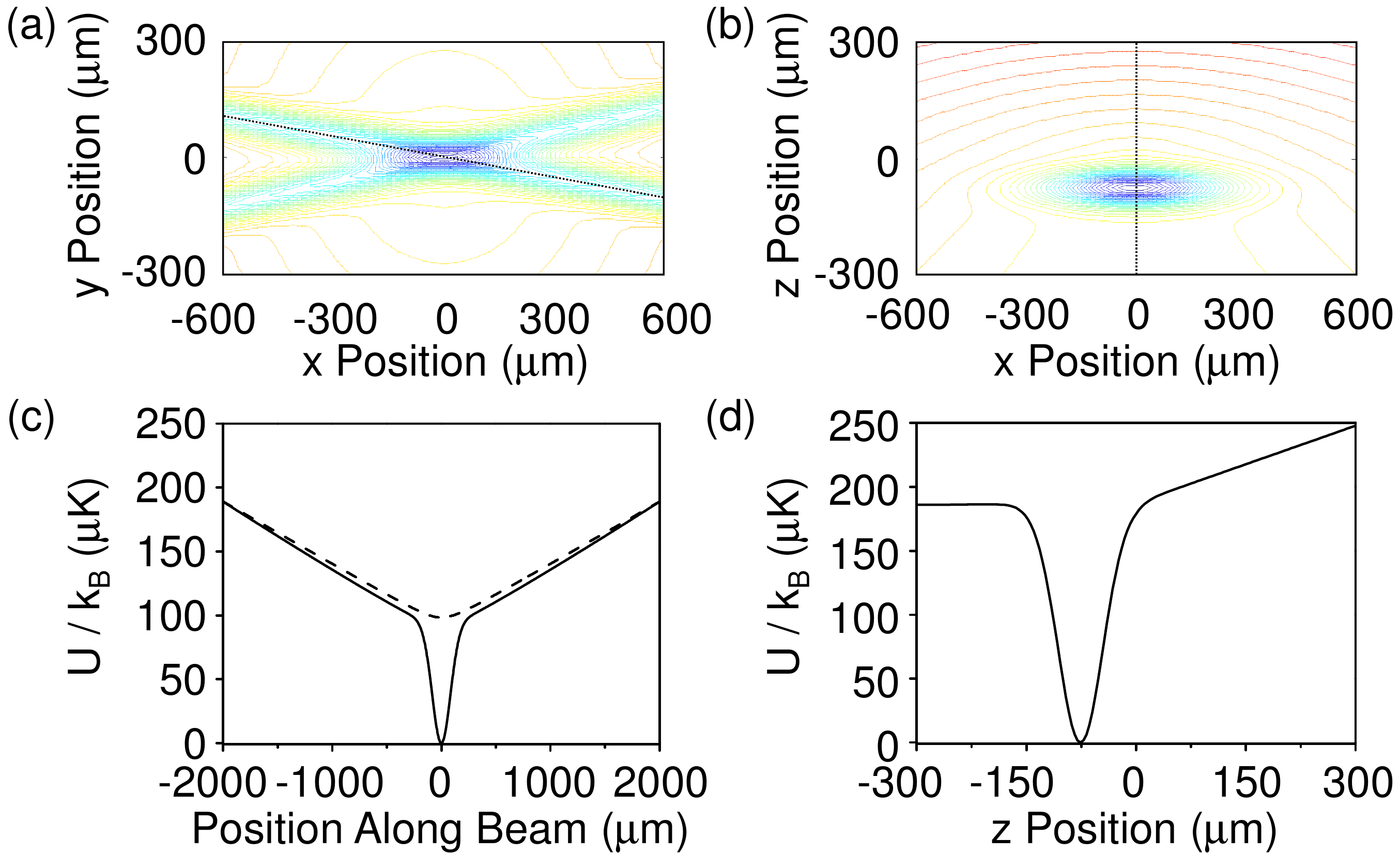}
\caption{The loading trap potential for \Rb\ atoms in the \abbmFminus\ state created with 6~W in each dipole beam and a magnetic field gradient of 29~G~cm$^{-1}$. Contour plots of the trap potential in (a) the $x-y$ plane intersecting the potential minimum and (b) the $x-z$ plane. Cross-sections through the potential minimum along one of the beams (c) and vertically (d). The crossed dipole trap is positioned $\sim80~\mu$m below the field zero of the quadrupole potential resulting in additional magnetic harmonic confinement along the beams. The purely magnetic contribution is shown as the dashed line in (c).}
\label{fig:LoadPot}%
\end{figure}

Atoms in the weak-field seeking $\left|1,-1\right\rangle$ state are loaded into the hybrid loading trap from the magnetic trap after pre-cooling using forced RF evaporation (see Section~\ref{chap:Expt} for more details). Initial loading is performed at a power of 6~W in each beam with a gradient of 29~G~cm$^{-1}$ and zero bias field. The magnetic field gradient is set slightly below the value which cancels gravity for the \abbmFminus\ state (\textit{i.e.} levitates it) and provides an attractive confining potential that adds to the existing optical confinement along the beam. The essential features of the trapping potential are presented in Figure~\ref{fig:LoadPot}. Along the beam the trap potential is the sum of the harmonic trap obtained from a quadrupole trap offset by a bias field and the dimple optical potential superimposed in the centre. In the horizontal plane, outside the dimple region, the trap potential increases linearly in all directions to values $>$ 1000~$\mu$K. The trap depth of 185~$\mu$K is thus entirely determined by the potential in the vertical $z$-direction, which is nearly constant on the lower side of the trap owing to the near cancellation of gravity. On the upper side of the trap a potential gradient of $\sim2g$ exists owing to the addition of magnetic and gravitational potentials. Hotter atoms are lost vertically downwards in the direction with the lowest trap depth. The frequencies of the loading trap are $\omega_{x}$~=~2$\pi \times$135~Hz and $\omega_{y} = \omega_{z}$~=~2$\pi \times$680~Hz.

\subsection{\label{sec:LevTrap}The levitated trap}

\begin{figure}%
\includegraphics[width=\columnwidth]{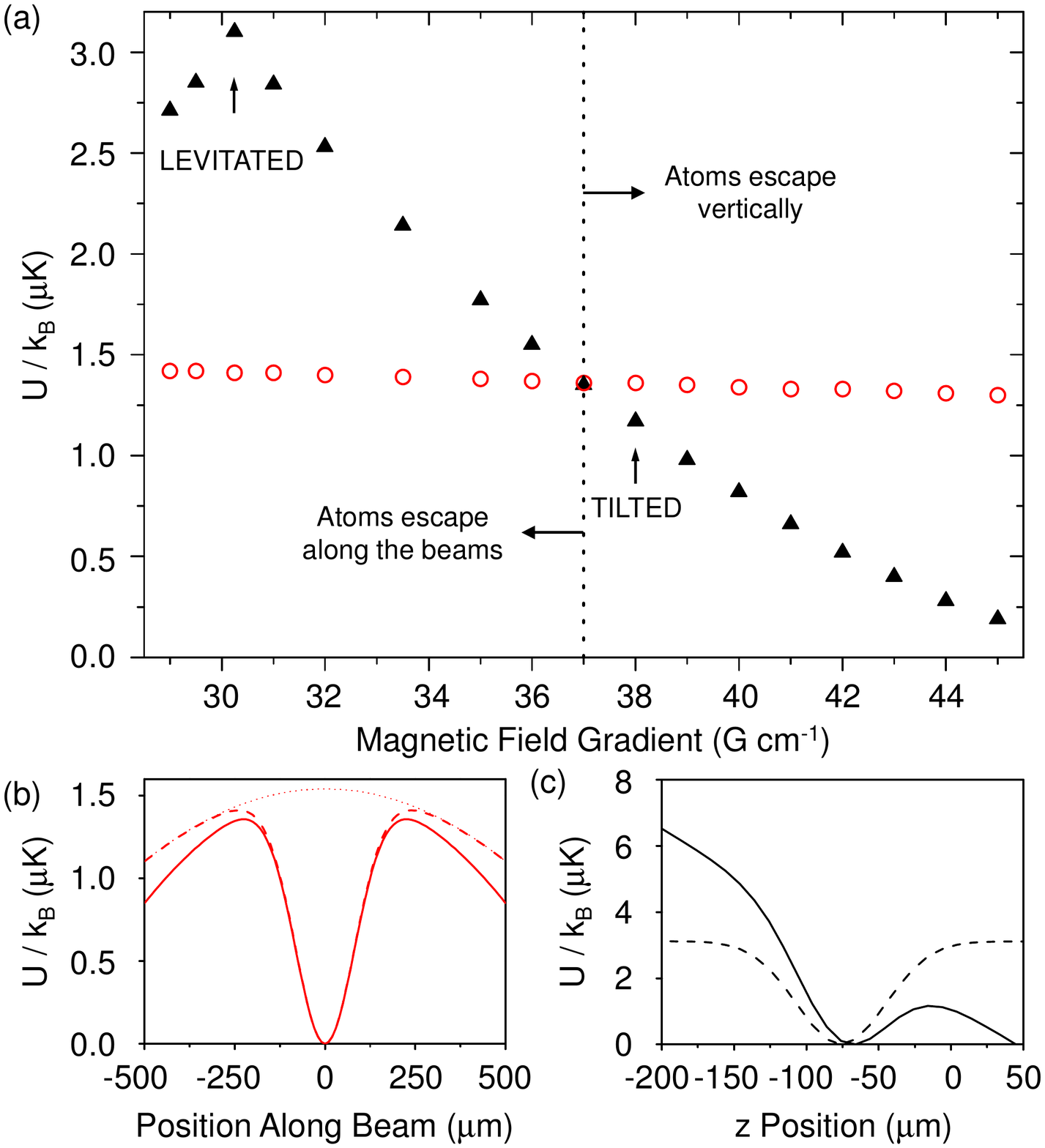}%
\caption{Controlling the depth of the levitated crossed dipole trap by the addition of a magnetic field gradient. (a) The trap depths for \Rb\ atoms in the \abbmFplus\ state in the $z$-direction (solid black triangles) and along the beams (open red circles) as a function of the magnetic field gradient. The beam powers are both 100~mW and the bias field is 22.4~G. Cross-sections through the potential minimum (b) along one of the beams and (c) vertically are shown for the levitated (dashed) and the tilted (solid) potentials indicated by the arrows in (a). The dotted line in (b) indicates the magnetic potential due to a gradient of 30.2~G~cm$^{-1}$ with no optical potential present.}
\label{fig:CrossDTPotential}
\end{figure}

The `levitated trap' differs from the loading trap in two main respects. Firstly a bias field of 22.4~G has been added which shifts the quadrupole zero downwards by $\sim7.5$~mm in order to levitate the atoms in the \abbmFplus\ state. Secondly the atoms are transferred into the high-field seeking $\left|1,+1\right\rangle$ state using rapid adiabatic passage (see Section~\ref{chap:Expt}). For this state the magnetic potential is anti-trapping and opposes the optical confinement. In this configuration, the nature of the evaporation surface of the hybrid trap depends in a useful way on the applied quadrupole gradient. This is shown in  Figure~\ref{fig:CrossDTPotential} which shows an example of how the horizontal and vertical trap depths vary with applied gradient for a fixed bias field of 22.4~G and with beam powers fixed at 100~mW. It can be seen that the trap depth along the beam (which is the shallowest part of the trap potential in the horizontal plane) is almost constant, whereas the vertical trap depth may be either greater or less than the horizontal depth. Thus the figure divides into two regions, with an approximate crossover gradient for this example of 37~G~cm$^{-1}$. The usefulness of this levitated trap stems from the existence of both regions. For gradients less than the crossover gradient evaporation occurs horizontally along the beams and can be achieved by decreasing the beam powers. This method also causes a consequent decrease in trap frequencies and hence the collision rate. This is generally detrimental to evaporative cooling although we note that it may be useful in some circumstances where three-body loss rates are high. For gradients greater than the crossover gradient evaporation occurs in the vertical direction owing to the lower trap depth (see Figure~\ref{fig:CrossDTPotential}~(c)). Physically this occurs because, as the gradient is increased, gravity is overcompensated and the residual upwards force (magnetic minus gravity) eventually overcomes the downwards confinement of the optical potential. The atoms escape upwards; this effect is referred to as `tilting' the trap \cite{Hung:AECOAIBECIOT} as it is analagous to tilting a cup full of water. In this region evaporation may be implemented by either reducing the beam power or by simply increasing the magnetic field gradient. In the latter case the trap frequencies do not change appreciably as the evaporation proceeds. 

Specific examples of the trap potentials are shown in Figure~\ref{fig:CrossDTPotential} (b) (horizontal cut) and Figure~\ref{fig:CrossDTPotential}~(c) (vertical cut) for two values of the gradient, one at 30.2~G~cm$^{-1}$, levitating the trap and the other at 38~G~cm$^{-1}$, tilting the trap. In Figure~\ref{fig:CrossDTPotential}~(a) it can be seen that changing the gradient has little effect on the horizontal trap depth whereas the vertical trap depth is highly dependent on the gradient. In both of these cases, changing the gradient has little effect on the trap frequencies. It has been observed in other work that when evaporating to degeneracy in \Cs\ the presence of a levitation field and tilting the trap can increase final condensate numbers \cite{Cho:TAQGOPRM}.

When perfectly aligned the untilted, levitated trap is almost symmetric about the potential minimum, as the magnetic and gravitational forces cancel either side of the dipole potential. However, misalignment of the beams such that the crossed dipole trap is no longer directly below the field zero of the quadrupole potential breaks the symmetry in the horizontal plane. The resulting offset of the magnetic and optical potentials means that the depth of the levitated trap is now lowest along just two of the four beams. This significantly reduces the surface where atoms can escape from the trap and is undesirable for evaporative cooling.

\section{\label{chap:Expt}Loading the dipole trap}

Full details of the vacuum system and MOT apparatus can be found in \cite{Harris:MTOACRBAM}. Here we briefly summarise the aspects of the apparatus employed to collect atoms. A pyramid MOT \cite{Arlt:APMOTAASOSA} is used as a cold atom source for a 6-beam MOT in a UHV glass cell. Up to 9(1)$\times$10$^{8}$ \Rb\ atoms can be loaded into the second MOT though we typically operate with 4(1)$\times$10$^{8}$ in order to preserve the vacuum. Following a compressed MOT and molasses stage the atoms are optically pumped into the \abbmFminus\ state and loaded into a magnetic quadrupole trap at an axial field gradient of 40~G~cm$^{-1}$. The gradient is adiabatically increased to 187~G~cm$^{-1}$ in order to increase the elastic collision rate. At this stage 1.7(1)$\times$10$^{8}$ 
 atoms remain at a temperature of 140(10)~$\mu$K and a PSD of 6.4(1)$\times$10$^{-7}$. The lifetime at this stage is 162(6)~s.

The atomic gas must be pre-cooled before it can be loaded into the optical trap as the initial temperature in the quadrupole trap is comparable to the $\sim180~\mu$K trap depth of the loading trap. Subsequently a large fraction of the cloud of atoms can be transferred from the magnetic reservoir into the optical potential. The pre-cooling is implemented using forced RF evaporation in the stiff magnetic trap at a gradient of 187~G~cm$^{-1}$. Two linear RF ramps are performed between frequencies of 30~MHz and 12~MHz in 12.5~s. After this stage, there remain 3.6(1)$\times$10$^{7}$ atoms at a temperature of 42(1)~$\mu$K and a PSD of 2.8(1)$\times$10$^{-5}$. At this temperature, the lifetime is reduced to 23(2)~s, limited by Majorana losses.

\begin{figure}%
\includegraphics[width=\columnwidth]{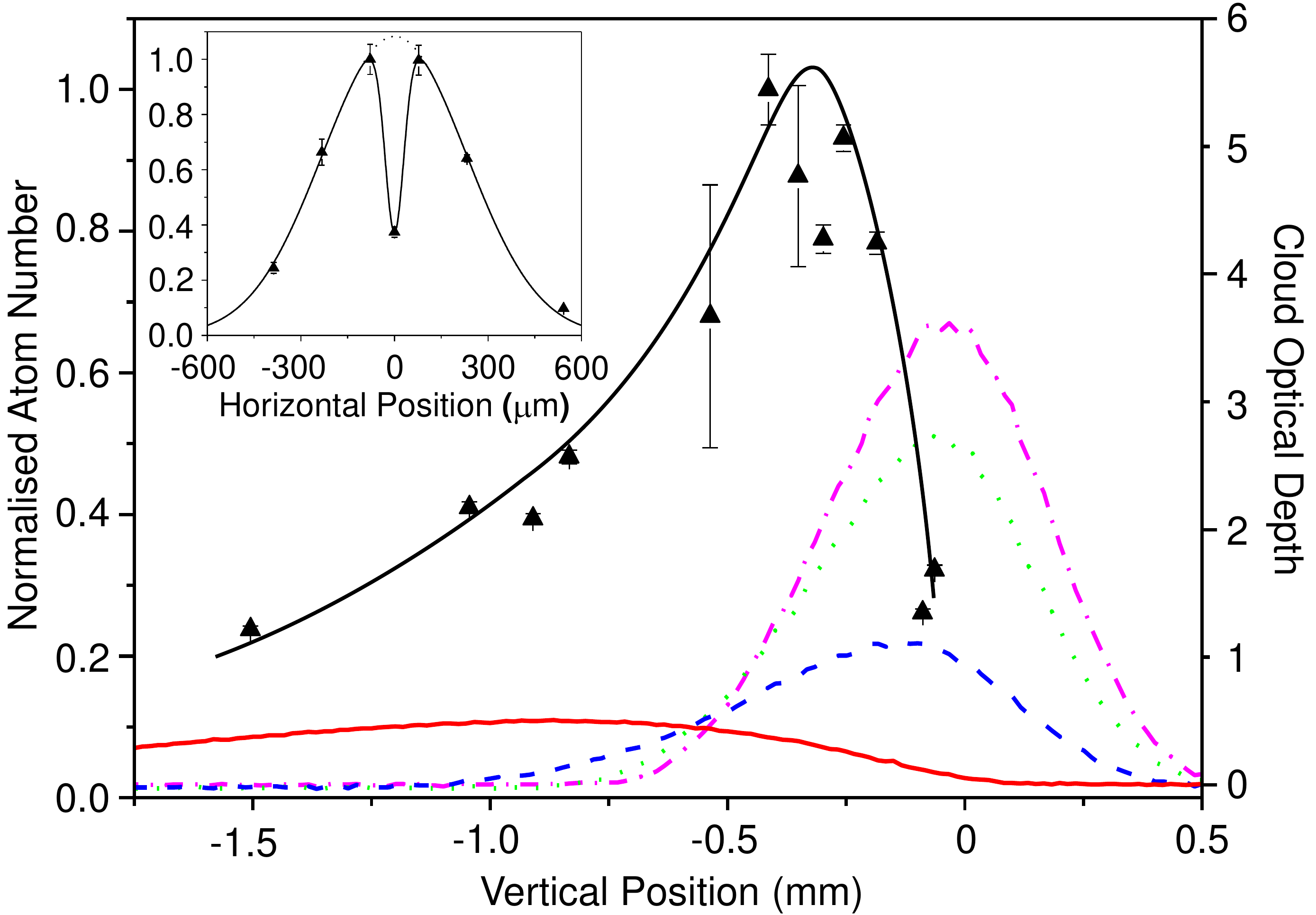}%
\caption{Loading the single beam dipole trap. Atom number (black triangles) is shown as a function of the vertical and horizontal (inset) beam position with respect to the magnetic field zero of the quadrupole potential. The beam power was 3.7~W, the waist was $\sim110~\mu$m and the gradient was 29~G~cm$^{-1}$. The solid and dotted black lines are to guide the eye only. The coloured lines show the evolution of the cloud density profile during loading as the magnetic field gradient is adiabatically ramped from the initial 187~G~cm$^{-1}$ (magenta dashed-dotted line) to the final 30~G~cm$^{-1}$ (red solid line). The data in the inset were taken for no vertical offset of the dipole trap and strikingly reveal the position of the field zero through the reduction in the trapped atom number due to Majorana spin flip losses.}%
\label{fig:CloudProfile}%
\end{figure}

Next, atoms are transferred to the hybrid loading trap described in Section~\ref{sec:LoadTrap}. To make the transfer, the magnetic field gradient and RF frequency are simultaneously decreased in 6~s to 29~G cm$^{-1}$ and 3~MHz respectively. During the ramp the optical potential is loaded through elastic collisions in a similar manner to the dimple trap \cite{Lin:RPO87RBECIACMAOP,Kraemer:OPOACBEC,Stamper-Kurn:RFOABEC}. The final gradient is slightly lower than the gradient required to levitate the atoms so that any atoms which are not confined in the optical trap fall out of the reservoir. The effective `evaporation' of these hotter atoms leads to a large increase of PSD in the optical trap. The coloured traces on Figure~\ref{fig:CloudProfile} show vertical profiles of the density of the atomic cloud as the magnetic field gradient is adiabatically decreased during the transfer. 

The alignment of the crossed dipole beams was optimised by exploiting the fact that each individual beam can form a complete hybrid trap when combined with the magnetic potential from the quadrupole trap. Using each beam to form such a `single beam trap' at separate times, we located the position at which the loading and evaporation were most efficient. Indeed, after this optimisation, we were able to make pure \Rb\ condensates of 2$\times$10$^{5}$ atoms in either single beam trap \cite{Lin:RPO87RBECIACMAOP}, which can be thought of as the ultimate diagnostic for evaporative cooling. In addition the single beams allow the intensity servo-mechanical circuits to be tested individually. To locate the optimum position for the beam in the $y$-direction the beam was fixed at the vertical position of the magnetic field zero (determined from absorption imaging of the magnetically trapped atoms). The beam was then scanned horizontally along the $y-$axis. A reduction in atom number was observed at the location of the field zero due to Majorana losses (see inset of Figure~\ref{fig:CloudProfile}).

To optimise the vertical position of each beam we investigated both the transfer efficiency and the subsequent evaporation efficiency as a function of the displacement from the quadrupole field zero. The radial trap frequency is largely independent of the vertical position as the radial confinement is provided by the beam, whereas the axial frequency is very sensitive to the vertical position as the axial confinement is provided by the magnetic trap. Although the dipole trap can be positioned either above or below the magnetic field zero, the combined trap is significantly deeper below the field zero and hence the beams are positioned there. Figure~\ref{fig:CloudProfile} displays the number of atoms loaded into a single beam dipole trap with beam waist $\sim110~\mu$m, as a function of position with respect to the magnetic field zero. The maximum number of atoms loaded into the dipole trap is observed approximately 3 beam waists below the magnetic field zero. However, our experiments with cooling to degeneracy in the single beam have revealed that the evaporation is more efficient when the beam is positioned approximately one beam waist below the field zero, in agreement with previous observations \cite{Lin:RPO87RBECIACMAOP}. This is due to the increase in trap frequency closer to the trap centre. For the current beam waists of $\sim60~\mu$m the optimum positions for the single beam traps were 80~$\mu$m below the field zero.

\begin{figure}%
\includegraphics[width=\columnwidth]{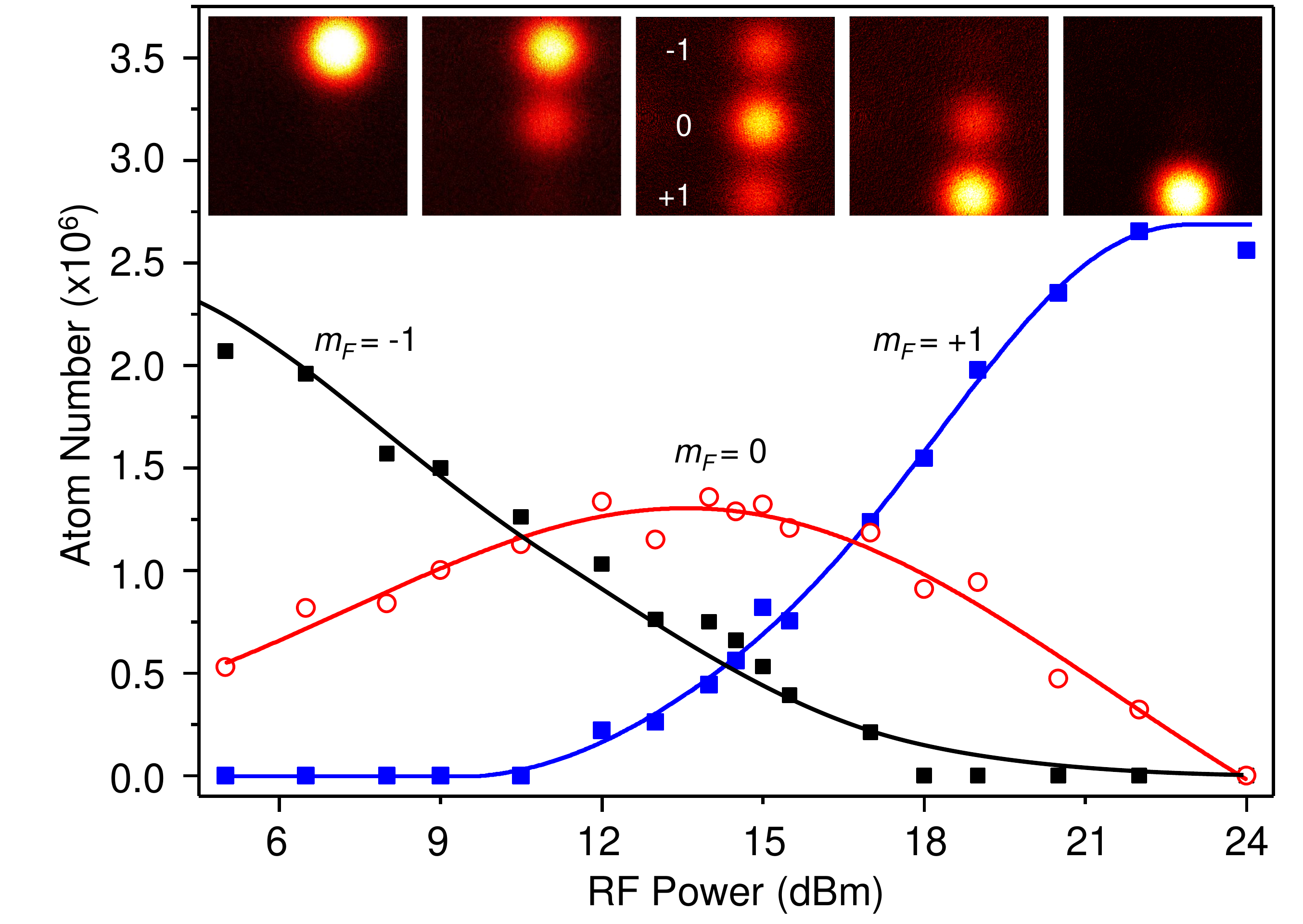}%
\caption{Spin transfer via rapid adiabatic passage. An RF field with a frequency of $\sim1.5~$MHz is applied to the atoms and the magnetic bias field is turned on to 22.8~G. A Stern-Gerlach measurement is used to analyse the spin composition of the gas as a function of the RF power. Typical absorption images indicate the efficient transfer of the population from the \abbmFminus\ state to the \abbmFplus\ state. Each image is $4.5\times4.5~$mm$^{2}$.}
\label{fig:RapidAdiabaticPassage}%
\end{figure}

After this initial transfer into the loading trap, we typically obtain 6.9(1)$\times$10$^{6}$ \Rb\ atoms in the \abbmFminus\ state at a temperature of 14(1)~$\mu$K.
\ The next step is to change to the levitated trap configuration; N.B. it is the \abbmFplus\ state that is levitated. All states are trappable in the crossed dipole trap and to transfer into the \abbmFplus\ state the technique of  rapid adiabatic passage is utilised \cite{Bergmann:CPTAQSOAAM}. To do this, an RF field at a frequency of 1.5~MHz is switched on by driving current through a 26$\times$26~mm$^{2}$ square coil of 3 turns placed a few cm from the trap. 
\ The bias field of 22.4~G is then switched on, which sweeps across the RF Zeeman resonance at 2.1~G as it ramps up in 18~ms to its final value.
\ The RF is then switched off. This process of RF sweeping adiabatically transfers the atoms into the \abbmFzero\ and/or \abbmFplus\ states depending on the amount of RF power applied to the coil. The transfer into the \abbmFplus\ state is performed with an RF power of 24~dBm, which gives a spin state purity of 96\%. 
 The different \mF\ states may be separated using Stern-Gerlach separation, by allowing the atoms to move freely in the magnetic field gradient for around 15~ms. Figure~\ref{fig:RapidAdiabaticPassage} displays the number of atoms present in each state as a function of RF power. The number in the \abbmFminus\ state decreases with increasing RF power, while the reverse is true for the \abbmFplus\ state. After this transfer 6.1(1)$\times$10$^{6}$ atoms with a temperature of 10.5(1)~$\mu$K and PSD of 1.6(1)$\times$10$^{-3}$ are present in the levitated crossed dipole trap in the \abbmFplus\ state.
\ The decrease in temperature follows from the slight reduction in the trap depth.
 
\section{\label{chap:Evap}Evaporation to BEC}

\begin{figure}%
\includegraphics[width=\columnwidth]{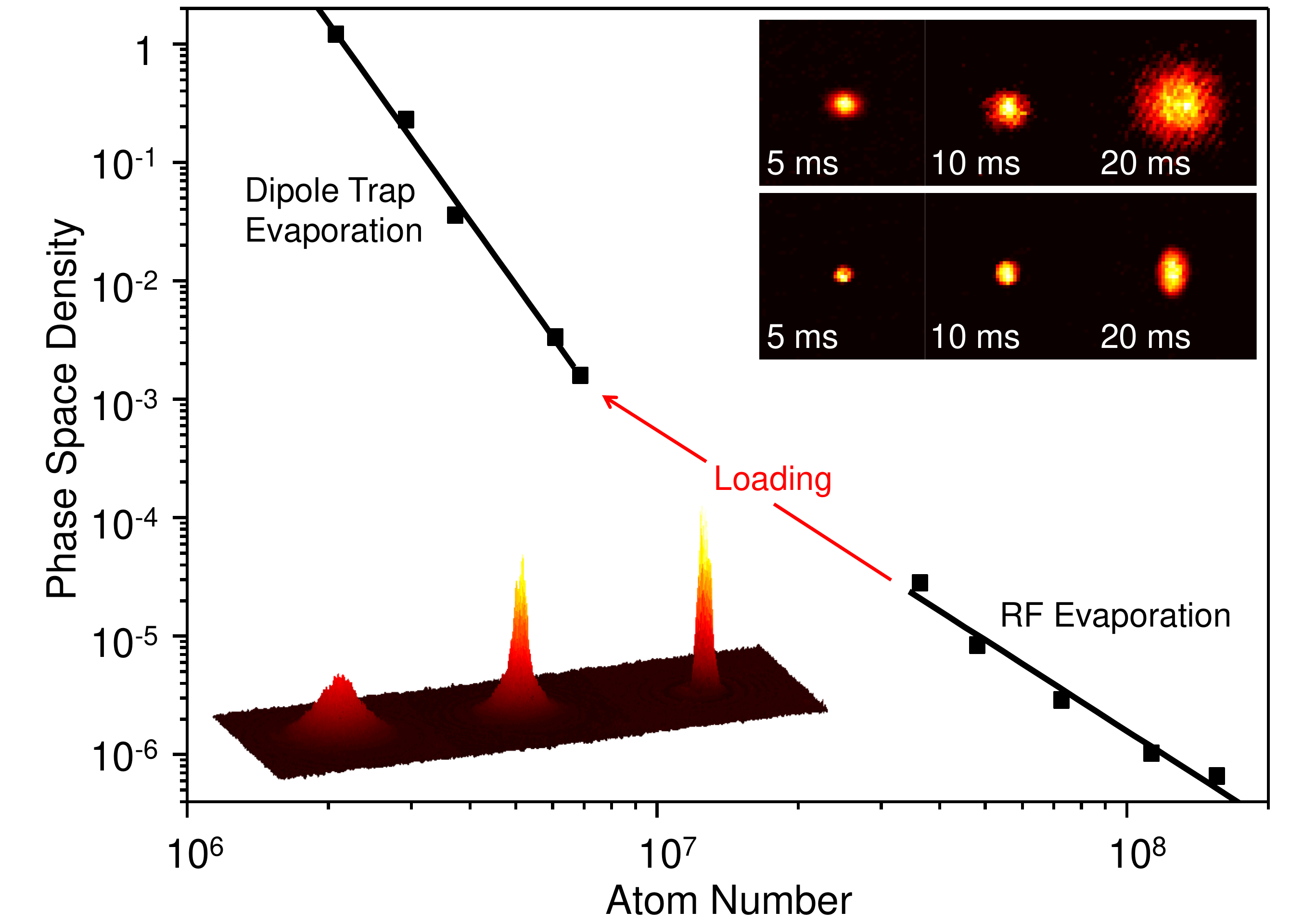}%
\caption{The phase-space density trajectory to BEC of \Rb\ atoms in the crossed dipole trap. The experimental sequence from right to left corresponds to RF evaporation in the quadrupole trap, dipole trap loading and forced evaporation in the dipole trap by reducing the beam powers. Characteristic signatures of the BEC transition are shown in absorption images, highlighting the anisotropic time of flight expansion of a condensate as compared to a thermal distribution (top right) and the evolution of the density distribution as the gas is cooled through the critical temperature (bottom left).}
\label{fig:PSDDiagram}%
\end{figure}

Evaporation of \Rb\ in the levitated dipole trap is performed by reducing the trap depth logarithmically in time. In practice we reduce the beam powers from the initial 6~W per beam to 150~mW per beam 
\ in 7.5~s. The trap depth in the final, shallow trap is 2.2~$\mu$K and the frequencies are $\omega_{x}$~=~2$\pi\times$21~Hz and $\omega_{y}$~=~$\omega_{z}$~=~2$\pi\times$110~Hz.
\ Despite the decrease in trapping frequencies as we reduce the beam powers, we found that we produced a \Rb\ condensate, without the necessity of a final tilting stage, although we expect this to be very important for \Cs\ (see Section~\ref{chap:Out}). At the end of the logarithmic ramp we observed the onset of BEC at a temperature of 0.20(1)~$\mu$K with $\sim7$\% of atoms in the condensate. The total atom number is 2.1(1)$\times$10$^{6}$. A pure condensate of $\sim1\times10^{6}$ atoms is obtained after holding in this final potential for 3~s. The PSD and number at the different stages of the evaporation trajectory are presented in Figure~\ref{fig:PSDDiagram}. This result demonstrates that our system acts as an excellent `refrigerator' for our refrigerant species \Rb.

\begin{figure}%
\includegraphics[width=\columnwidth]{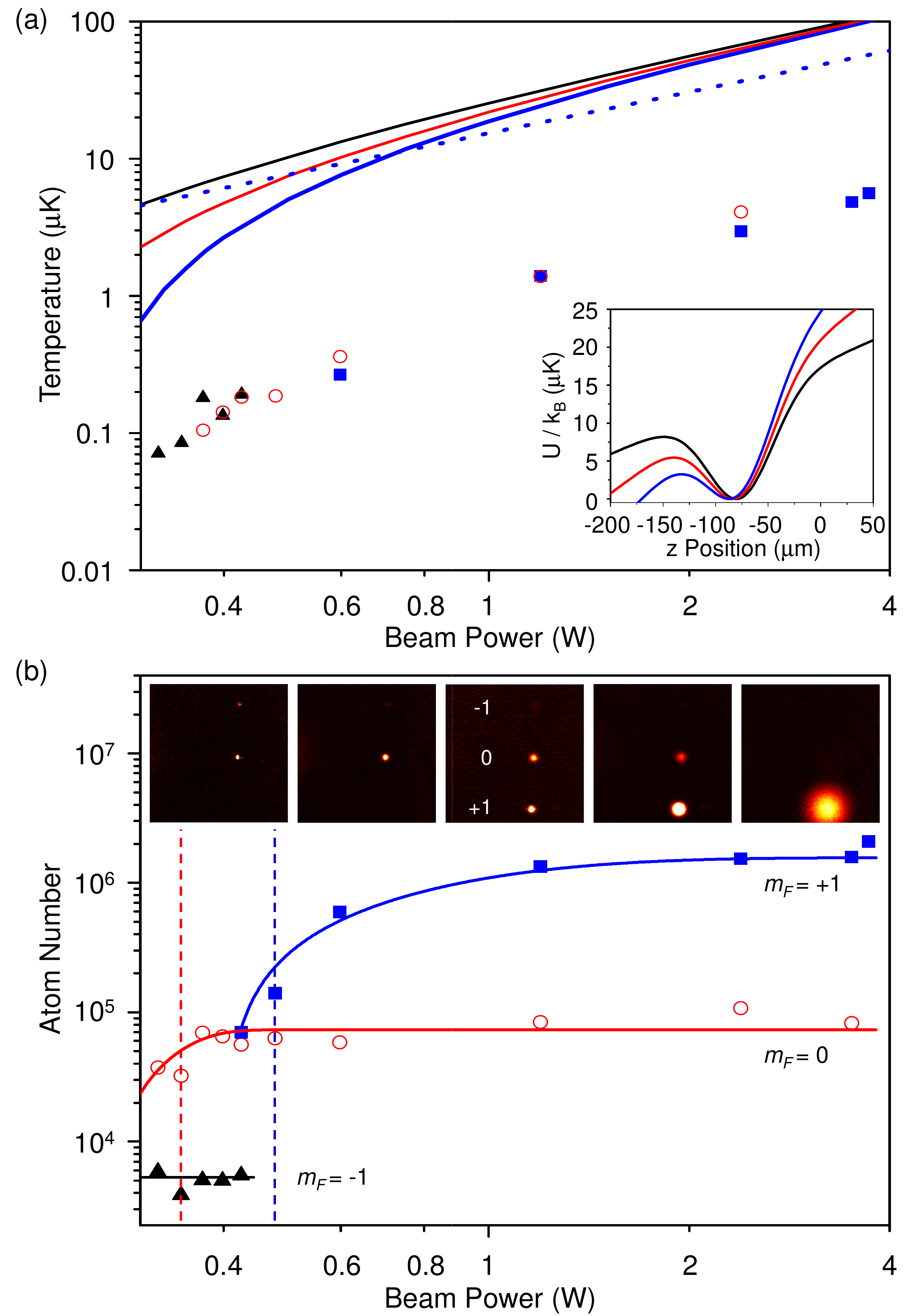}%
\caption{Sympathetic cooling of different spin states in the crossed dipole trap. (a) Temperatures and trap depths for each \mF\ state and (b) atom number as a function of the power in each beam. Data are shown for the \abbmFplus\ state (blue lines and squares), the \abbmFzero\ state (red lines and circles) and the \abbmFminus\ state (black lines and triangles). The solid lines in (a) indicate the trap depth along $z$, the dashed line indicates the trap depth along the beam and allows identification of the beam powers at which the evaporation surface switches from horizontal to vertical (tilting): these beam powers are 0.8~W and 0.45~W for \abbmFplus\ and \abbmFzero\ respectively. The inset shows the vertical cross-section through the potential minimum for a beam power of 0.45~W and a magnetic field gradient of 13~G~cm$^{-1}$. The solid lines in (b) are to guide the eye only and the dashed lines indicate the onset of Bose-Einstein condensation. The absorption images in (b) are taken after a Stern-Gerlach time of flight and highlight the sympathetic cooling of the \abbmFzero\ state.}%
\label{fig:SympCool}%
\end{figure}

In a separate experiment we deliberately load a mixture of \mF\ states into the levitated crossed dipole trap in order to demonstrate sympathetic cooling of the \abbmFminus\ and \abbmFzero\ states by a refrigerant of \abbmFplus\ atoms. The trap was loaded with 3.7~W in each beam and after the RF adiabatic transfer, of the remaining trapped atoms, 95.8\% are in the \abbmFplus\ state, 3.8\% of atoms are in the \abbmFzero\ state, and 0.4\% are in the \abbmFminus\ state.
\ The trap depth due to the purely optical potential is equal for each \mF\ state however for this experiment we add a magnetic field gradient of 13~G~cm$^{-1}$ and a bias field of -22.8~G (reversed sign so that it shifts the field zero upwards). This modifies the trap as follows. The offset quadrupole field now only partially levitates the \abbmFminus\ atoms whilst the \abbmFplus\ atoms experience a downwards force equivalent to an acceleration of $\approx1.5~g$ and this modifies the trap depth for each state. The gradient of 13~G~cm$^{-1}$ has been chosen to allow tilting of the trap for the \abbmFplus\ atoms. This is illustrated in Figure~\ref{fig:SympCool}~(a) which shows how the trap depths for each \mF\ state vary as a function of the power in each beam. The dotted line shows the trap depth for the \abbmFplus\ state along the beam. The ratio of trap depths in the \abbmFplus\, \abbmFzero\ and \abbmFminus\ states varies from 1.0~:~1.1~:~1.9 to 1.0~:~1.6~:~2.3 for beam powers of 3.7~W and 0.45~W respectively. This means that when the beam powers are reduced the hotter and more numerous  \abbmFplus\ atoms will be evaporated first, and the ensemble will then cool via rethermalisation. This sympathetically cools the other two hyperfine states with little loss in atom number. As the beam power is reduced below 0.8~W, the evaporation surface changes from horizontally along the beam to a tilting evaporation in the $-z$-direction. As more \abbmFplus\ atoms are removed the temperature of the other hyperfine states begins to decrease, as indicated in Figure~\ref{fig:SympCool}~(a). The corresponding changes in number are plotted in Figure~\ref{fig:SympCool}~(b) where the false colour absorption images show how the atoms in the less populated states become more visible as the cloud cools and optical depth increases.

The \abbmFplus\ atoms form a pure BEC of 1.4$\times$10$^{5}$ atoms at a beam power of 0.48~W.
\ All \abbmFplus\ atoms are lost at a beam power of $\sim0.4$~W. At this beam power the trap depth of the \abbmFzero\ state is 4.8~$\mu$K, and 7.4~$\mu$K for the \abbmFminus\ state. The role of the \abbmFzero\ atoms is now to sympathetically cool the \abbmFminus\ atoms, again via tilting evaporation as the beam power is reduced towards 0.3~W. A BEC is produced in the \abbmFzero\ state containing 3$\times$10$^{4}$ atoms, where the temperature of the thermal \abbmFminus\ state is 90(2)~nK. Plain evaporation of the remaining atoms in the \abbmFminus\ state leads to further cooling of this state, close to degeneracy with PSD of $\sim0.5$, after which there are too few atoms to image reliably.

\section{\label{chap:Out}Outlook: Sympathetic cooling of Cs}

We have reported an apparatus and method capable of efficiently cooling a large number of \Rb\ atoms to the BEC transition. As such we have constructed an excellent `refrigerator' and now discuss the prospect of using the \Rb\ refrigerant for sympathetic cooling of \Cs\ in the $\left|3,+3\right\rangle$ state.

It is known that \Cs\ possesses a narrow window for BEC in the absolute ground state $\left|3,+3\right\rangle$ for bias fields in the range of $21-25$~G \cite{Weber:BECOC,Weber:TBRALSLIAUAG}. Within this window the ratio of elastic to three-body collisions is sufficiently high for successful evaporative cooling. However, the \Cs\ $\left|3,+3\right\rangle$ state is high-field seeking and hence must be confined in a dipole or hybrid trap where the magnetic field is a free parameter. It is also desirable to have flexible, independent control over the trap depths and frequencies in order to maintain densities that suppress three-body losses whilst carrying out forced evaporation.

\begin{figure}%
\includegraphics[width=\columnwidth]{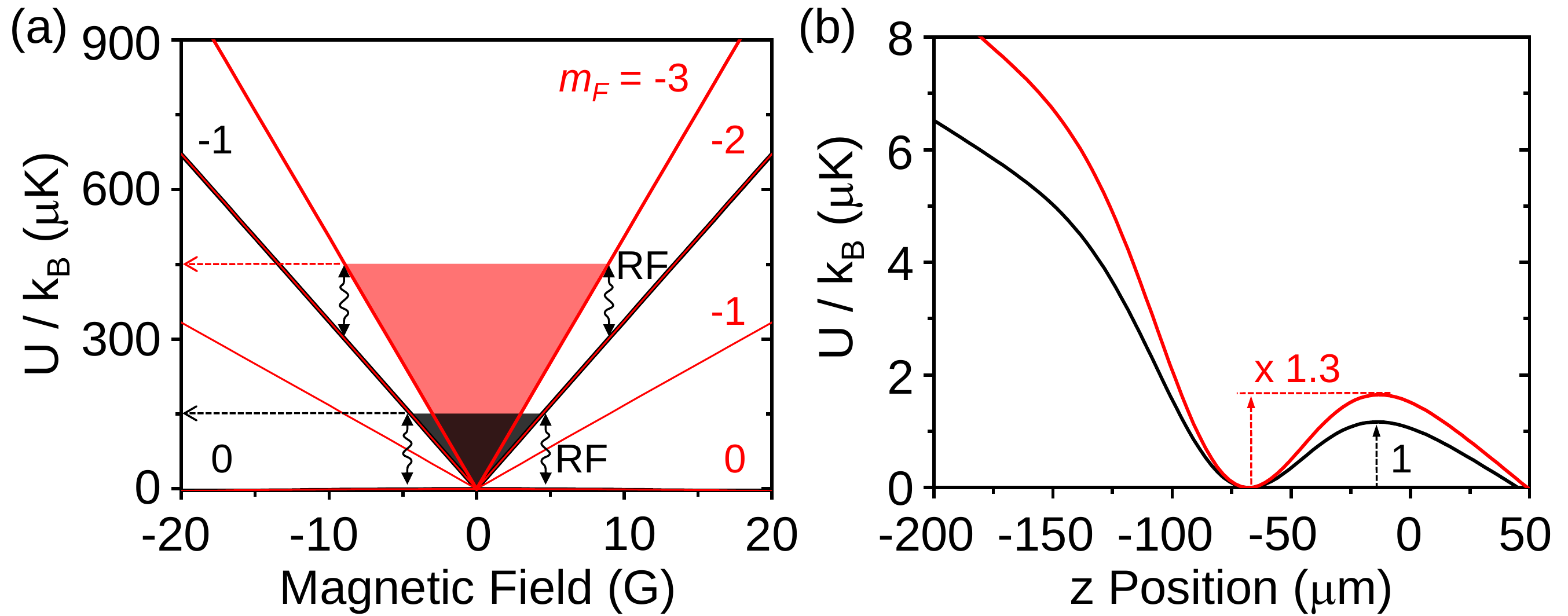}
\caption{Application of the experiment to sympathetic cooling of \Cs. (a) The magnetic quadrupole and (b) the tilted crossed dipole trap potentials for \Rb\ (black) and \Cs\ (red). The depth of the quadrupole trap determined by the RF frequency is $3\times$ larger for \Cs\ than for \Rb, thereby permitting sympathetic cooling. Similarly the dipole trap potential is deeper for \Cs\ than for \Rb\ due to the greater polarisability at 1550~nm. The example potential in (b) corresponds to 100~mW in each beam, a magnetic field gradient of 38~G~cm$^{-1}$ and a bias field of 22.8~G.}
\label{fig:Outlook}%
\end{figure}

Our hybrid system is ideal for achieving these requirements for a number of reasons related to the complimentary atomic properties of \Cs\ and \Rb. Firstly, the magnetic moment to mass ratios of \Cs\ and \Rb\ in their absolute internal ground states differ by $<2$\%. This leads to excellent spatial overlap in a magnetic trap and also means that the two species can be effectively levitated with the same magnetic field gradient. Secondly, throughout the evaporative cooling stages of our experiment the trap depth is greater for \Cs\ than \Rb, meaning the \Rb\ refrigerant will always be preferentially removed whilst the \Cs\ target will be cooled sympathetically. In the magnetic trap the depth determined by the RF frequency is exactly $3\times$ larger for \Cs\ than for \Rb\ (see Figure~\ref{fig:Outlook}~(a)). Similarly the optical potential is $\sim1.3\times$ deeper for \Cs\ than for \Rb\ due to the greater polarisability at 1550~nm. The calculated polarisabilities in atomic units are $\sim572$~a$_{0}^{3}$ for \Cs\ and $\sim425$~a$_{0}^{3}$ for \Rb\ (where a$_{0}=0.0529$~nm) \cite{Safranova:FDPOAMAFUTISR}. We note that the calculated ratio of the trap depths is very similar to that of the different \mF\ states discussed in Section~\ref{chap:Evap} where we showed the sympathetic cooling to be highly effective. Finally in our hybrid trap the trap `tilting' technique can be effectively employed in the final stages of the evaporation, again to selectively evaporate the refrigerant \Rb\ atoms (see Figure~\ref{fig:Outlook}~(b)).

Whilst our apparatus and method are ideally suited to sympathetic cooling of \Cs\ by \Rb\, the success of the approach will ultimately depend on the interspecies elastic and inelastic collision cross sections. At present, the interspecies scattering length and its dependence on magnetic field is not fully understood, although a definite picture is emerging. Early work on sympathetically cooling \Cs\ by \Rb\ focussed on the magnetically trappable states \cite{Anderlini:SCACPOARBM,Haas:SSMCOAMORACA}. Analysis of the rethermalisation rates suggested a large interspecies triplet scattering length \cite{Tiesinga:DOTSLOTA3SPPORC}. More recently, numerous interspecies Feshbach resonances have been reported \cite{Pilch:OOIFRIAURCM}. The interpretation of this Feshbach spectrum is ongoing; however preliminary findings indicate a bound state close to threshold, implying a large interspecies scattering length \cite{Hutson:PC}. In fact, recent observations using magnetic field modulation spectroscopy have detected a weakly bound Feshbach level close to threshold \cite{Lercher:POADSBECORACA}. A large interspecies scattering length would imply a large interspecies three-body loss rate coefficient through the $a^{4}$ scaling \cite{Braaten:EPICA}. Such high losses have been reported \cite{Cho:TAQGOPRM,Lercher:POADSBECORACA}. However, the different density scaling of two-body elastic and three-body inelastic collisions should allow a trap regime to be devised which still gives favourable conditions for sympathetic cooling. Indeed, we present our experimental demonstration of efficient sympathetic cooling elsewhere \cite{Cho:TAQGOPRM,McCarron:IIAQDMORAC}.

In conclusion, we have described in detail an apparatus and method designed for sympathetic cooling of \Cs\ by \Rb\ in a levitated crossed dipole trap. The method combines the large recapture of a magnetic quadrupole trap and efficient RF evaporation therein prior to loading the atoms into the optical trap. A simple approach for optimising the transfer into the dipole trap that exploits the axial magnetic confinement in the hybrid magnetic-optical trap has been described. The production of BECs of $\sim1\times$10$^{6}$ atoms demonstrates the overall success of the approach for efficiently cooling the \Rb\ refrigerant. The application of the approach to the cooling of atomic mixtures of \Rb\ and \Cs\ is presented elsewhere in this special issue \cite{Cho:TAQGOPRM}.

We acknowledge financial support from the EPSRC (GR/S78339/01, EP/E041604/1, EP/H03363/1) and the European Science Foundation within the framework of the EuroQUAM collaborative research project QuDipMol. SLC acknowledges the support of the Royal Society.

\bibliographystyle{epj}

\end{document}